# Evaluating the Security of Merkle Trees in the Internet of Things: An Analysis of Data Falsification Probabilities

Oleksandr Kuznetsov, *Member, IEEE*, Alex Rusnak, Anton Yezhov, Kateryna Kuznetsova, Dzianis Kanonik, and Oleksandr Domin

*Abstract*—Addressing the critical challenge of ensuring data integrity in decentralized systems, this paper delves into the underexplored area of data falsification probabilities within Merkle Trees, which are pivotal in blockchain and Internet of Things (IoT) technologies. Despite their widespread use, a comprehensive understanding of the probabilistic aspects of data security in these structures remains a gap in current research. Our study aims to bridge this gap by developing a theoretical framework to calculate the probability of data falsification, taking into account various scenarios based on the length of the Merkle path and hash length. The research progresses from the derivation of an exact formula for falsification probability to an approximation suitable for cases with significantly large hash lengths. Empirical experiments validate the theoretical models, exploring simulations with diverse hash lengths and Merkle path lengths. The findings reveal a decrease in falsification probability with increasing hash length and an inverse relationship with longer Merkle paths. A numerical analysis quantifies the discrepancy between exact and approximate probabilities, underscoring the conditions for the effective application of the approximation. This work offers crucial insights into optimizing Merkle Tree structures for bolstering security in blockchain and IoT systems, achieving a balance between computational efficiency and data integrity.

## I. INTRODUCTION

In the rapidly evolving landscape of the Internet of Things (IoT), ensuring the integrity and authenticity of data has become a paramount concern [1]. The proliferation of interconnected devices has ushered in an era where data is continuously generated, processed, and exchanged, making systems vulnerable to various forms of cyberattacks [2]. One of the most significant challenges in this domain is safeguarding against data tampering and ensuring the reliability of information exchanged between devices [3]. In this context, the application of Merkle trees, a fundamental component in blockchain technology known for ensuring data integrity, presents a novel approach to enhancing IoT security [4], [5].

This paper aims to delve into the analysis of Merkle trees' security within the IoT paradigm. Specifically, it focuses on the critical evaluation of data falsification probabilities under a fixed Merkle path scenario. Such an analysis is crucial, as root collisions in Merkle trees could potentially allow malicious actors to forge data, thereby deceiving the system and compromising its reliability. While Merkle trees are renowned for their use in blockchain systems to prove data integrity and authenticity, their application in the IoT domain necessitates a thorough examination of their effectiveness and potential vulnerabilities.

The study presents both exact mathematical formulations and approximate estimations to assess the likelihood of these root collisions. The implications of these findings are significant for blockchain systems, where Merkle trees are extensively employed. However, our focus extends beyond the blockchain, venturing into the realm of IoT, where the security implications are profound and the context differs markedly. By exploring the nuances of Merkle trees in this new environment, this paper aims to contribute to the broader discourse on IoT security, offering insights and recommendations for leveraging Merkle trees to fortify IoT networks against data tampering and ensure the veracity of the information within these interconnected systems.

In essence, this paper endeavors to bridge the gap between the established utility of Merkle trees in blockchain technology and their prospective application in securing IoT ecosystems. Through rigorous analysis and assessment, it seeks to illuminate the potential and limitations of Merkle trees in a domain where data integrity is not just a necessity but a cornerstone of operational reliability and trust.

Manuscript received January 3, 2024.
This work was supported by Proxima Labs, 1501 Larkin Street, suite 300, San Francisco, USA.
*(Corresponding author: Oleksandr Kuznetsov.)*
All authors contributed equally to this work.

Oleksandr Kuznetsov (Member, IEEE) is with the Department of Political Sciences, Communication and International Relations, University of Macerata, Via Crescimbeni, 30/32, 62100 Macerata, Italy. He is also with the Proxima Labs, 1501 Larkin Street, suite 300, San Francisco, USA (e-mails: kuznetsov@karazin.ua, oleksandr.k@zpoken.io).
Alex Rusnak is with the Proxima Labs, 1501 Larkin Street, suite 300, San Francisco, USA (e-mail: alex@proxima.one).
Anton Yezhov is with the Proxima Labs, 1501 Larkin Street, suite 300, San Francisco, USA (e-mail: anton.yezhov@zpoken.io).
Kateryna Kuznetsova is with the Proxima Labs, 1501 Larkin Street, suite 300, San Francisco, USA (e-mail: kateryna.k@zpoken.io).
Dzianis Kanonik is with the Proxima Labs, 1501 Larkin Street, suite 300, San Francisco, USA (e-mail: denis@proxima.one).
Oleksandr Domin is with the Proxima Labs, 1501 Larkin Street, suite 300, San Francisco, USA (e-mail: scmaster@zpoken.io).

## II. Literature Review

Freitag et al. [6] explore the time-space tradeoffs in sponge hashing, offering a novel perspective on collision resistance. Their study on short collisions within this context is particularly noteworthy for its insights into the sponge construction's parameters. This work is instrumental in understanding the limitations and attack vulnerabilities of sponge hashing, thereby contributing significantly to the field of cryptographic hashing methods.

In another pivotal study [7], Ghoshal and Komargodski examine preprocessing adversaries in Merkle-Damgård (MD) hashing. Their focus on bounded-length collisions in the random oracle model provides a nuanced understanding of the MD hashing's resilience against certain types of attacks. This research is particularly relevant in evaluating the security of widely-used hashing techniques in the context of IoT devices.

Hu and colleagues delve into the vulnerabilities of the Merkle approach in proving liabilities [8]. Their analysis of the Maxwell protocol and its susceptibility to underreporting liabilities offers critical insights into the weaknesses of Merkle-based systems in decentralized environments. This research is particularly relevant for understanding the security of transaction systems like Bitcoin exchanges and has profound implications for blockchain technology.

The work by Kumari et al. [9] introduces the Signature-based Merkle Hash Multiplication (SMHM) algorithm, aimed at securing communication in IoT devices. Their approach, considering the challenges posed by 6G technology, offers a forward-looking perspective on securing IoT networks against quantum computing threats. This research contributes significantly to developing robust security protocols in the IoT realm.

Mitra and team [10] propose the Polar Coded Merkle Tree (PCMT) to improve the detection of Data Availability (DA) attacks in blockchain systems. Their approach addresses the limitations of previous coding techniques and presents a viable solution for large blockchains. This innovation is crucial for enhancing the security and reliability of blockchain systems against DA attacks.

In a subsequent study, the same authors introduce the Graph Coded Merkle Tree (GCMT), further refining their approach to mitigating DA attacks in blockchain systems [11]. Their informed design of polar factor graphs and the focus on large block size applications provide a comprehensive solution to previously identified challenges, marking a significant advancement in blockchain security.

Rao et al. [12] propose a dynamic outsourced auditing scheme for cloud storage, utilizing a batch-leaves-authenticated Merkle Hash Tree. Their approach addresses the trust issues in third-party auditing and supports verifiable dynamic updates, which is critical for cloud computing security. This research fills a vital gap in ensuring data integrity in outsourced environments.

Sarkar's study [13] introduces a new domain extender for collision-resistant hash functions, improving upon the Merkle–Damgård iteration. The proposed directed acyclic graph-based construction offers a more efficient alternative for hashing arbitrary length strings. This innovation significantly reduces computational requirements, paving the way for more efficient hashing methods.

Xu and colleagues [14] develop a dynamic Fully Homomorphic encryption-based Merkle Tree (FHMT) for lightweight streaming authenticated data structures. Their work is particularly relevant for streaming data environments, offering a balanced performance between client and server. This research is pivotal in advancing the use of Merkle trees in dynamic, resource-limited contexts.

Zhu et al. [4] propose an improved convolution Merkle tree-based blockchain scheme for secure electronic medical record storage. Their innovative approach in employing a convolutional layer structure significantly enhances efficiency and security, making it a notable contribution to the field of secure data storage and transmission, especially in healthcare.

While these studies collectively advance our understanding of Merkle trees in various contexts, including blockchain, IoT, and cloud computing, there is a noticeable gap in the literature specifically addressing the application of Merkle trees in the context of IoT security. Our research aims to fill this gap by providing a comprehensive analysis of Merkle tree root collisions and data falsification within the IoT framework, thereby offering new insights and solutions for enhancing IoT data integrity and security.

## III. Background

In the burgeoning realms of the Internet of Things (IoT) and blockchain technology, maintaining the integrity and authenticity of data is a pivotal challenge. The IoT ecosystem is characterized by a vast array of interconnected devices, continuously generating, processing, and exchanging data. This environment, inherently diverse and dynamic, poses significant risks concerning data security, particularly in aspects of tampering and authenticity. Similarly, blockchain technology, while renowned for its robust security mechanisms, confronts challenges in ensuring the immutability and verification of the vast amounts of data processed in its networks.

### A. Merkle Trees: Mechanism and Application

Merkle trees, conceptualized by Ralph Merkle, present an efficient and secure method to verify the content of large data structures. A Merkle tree is a binary tree in which every leaf node contains the hash of a data block, and every non-leaf node contains the cryptographic hash of its child nodes.

The construction and functionality of Merkle Trees are predicated on cryptographic principles, ensuring data integrity and authenticity through a series of mathematical operations. A Merkle Tree is built from the bottom up, starting with a set of data blocks (Figure 1).

Let $D = D_1, D_2, \ldots, D_n$ be the set of data blocks. Each block $D_i$ is subjected to a cryptographic hash function $H$, generating a set of leaf nodes $L = L_1, L_2, \ldots, L_n$, where

$L_i = H(D_i)$.

If $n$ is odd, an additional leaf node duplicating the last hash is added to maintain a balanced tree structure. The internal nodes are then constructed by recursively hashing pairs of child nodes:

$$N_{ij} = H(N_i \| N_j),$$

where $N_i$ and $N_j$ are child nodes, and $N_{ij}$ is their parent node. This process is iteratively applied until the root node, or Merkle Root $R$, is derived, representing the entire data set's hash:

$$R = H(N_{root_{left}} \| N_{root_{right}}).$$

The path in a Merkle Tree refers to the sequence of nodes and hashes required to verify a particular data block's integrity. For a given block $D_i$, its path $\mathcal{P}(D_i)$ is the set of sibling nodes and their parent hashes leading up to the root. Formally:

$$\mathcal{P}(D_i) = \{N_{sib_1}, N_{sib_2}, ..., N_{sib_m}\},$$

where $N_{sib_j}$ is the sibling node and $N_{par_{j-1}}$ is the parent node, so we have:

$$L_i = H(D_i); N_{par_0} = L_i;$$
$$N_{par_1} = H(N_{par_0} \| N_{sib_1});$$
$$N_{par_2} = H(N_{par_1} \| N_{sib_2});$$
$$\ldots$$
$$R = N_{par_m} = H(N_{par_{m-1}} \| N_{sib_m}).$$

To verify the integrity and authenticity of a data block $D_i$, one must reconstruct the path from $L_i$ to the root $R$ and compare it with the known Merkle Root. The verification process involves the following steps:

1. Initial Hashing: Compute the hash of the data block:
$$L_i = H(D_i); N_{par_0} = L_i.$$

2. Path Reconstruction: For each $N_{sib_j}$ in $\mathcal{P}(D_i)$, compute the parent hash:
$$N_{par_j} = H(N_{par_{j-1}} \| N_{sib_j}).$$

3. Root Comparison: Ascend the tree, iteratively applying the hash function until the reconstructed root $R'$ is obtained. The data block $D_i$ is authentic and unaltered if and only if:
$$R' = R.$$

This process ensures that any alteration in $D_i$ or its path would result in a different $R'$, thereby detecting tampering or data corruption.

*B. Application in IoT and Blockchain*

In the context of IoT, Merkle trees offer a method to ensure the integrity and authenticity of data across diverse and distributed devices. They enable IoT systems to efficiently validate data integrity without the need for transmitting large volumes of data, thereby reducing bandwidth and processing requirements. This is particularly beneficial in scenarios where IoT devices have limited computational and storage capabilities.

In blockchain systems, Merkle trees are integral to the construction of blocks. Each transaction in a block is represented as a leaf node in a Merkle tree, with the root hash being included in the block's header. This mechanism ensures that any alteration in a transaction would result in a different block header, thereby maintaining the blockchain's integrity. Furthermore, Merkle trees enable light clients in blockchain networks to efficiently verify the existence and integrity of transactions without downloading the entire blockchain.

Thus, Merkle trees serve as a cornerstone in ensuring data integrity and authenticity in both IoT and blockchain systems. Their application effectively addresses the challenges posed by the vast, distributed nature of these systems, providing a scalable and secure solution for data verification. This background forms the basis for our exploration into the specific application of Merkle trees in assessing root collision and data falsification probabilities within the IoT framework, a critical aspect for advancing the security and reliability of these emerging technologies.

*C. Problem Statement*

In the context of blockchain networks and the Internet of Things (IoT), ensuring the integrity and authenticity of data is paramount. A critical concern is the potential for data substitution within Merkle Trees, where altered data $D_i'$ might inadvertently or maliciously replace the original data $D_i$ without detection. Given a Merkle Tree with a set of data blocks $D_1, D_2, \ldots, D_n$, we aim to evaluate the probability $P(R = R')$ that the Merkle Root $R$ of the original tree remains unchanged when a specific data block $D_i$ is substituted with $D_i'$, while all other data blocks remain constant. Formally, we define:

$$P_{\text{falsification}} = P(R = R') = P(N_{par_m} = N_{par_m}'), \quad (1)$$

where
$$N_{par_m} = H(N_{par_{m-1}} \| N_{sib_m}) =$$
$$= H(H(N_{par_{m-2}} \| N_{sib_{m-1}}) \| N_{sib_m}) =$$
$$\ldots$$
$$= H(H(...H(N_{par_0} \| N_{sib_1})... \| N_{sib_{m-1}}) \| N_{sib_m}) =$$
$$= H(H(...H(H(D_i) \| N_{sib_1})... \| N_{sib_{m-1}}) \| N_{sib_m})$$

and

$$N_{par_m}' = H(N_{par_{m-1}}' \| N_{sib_m}) =$$
$$= H(H(N_{par_{m-2}}' \| N_{sib_{m-1}}) \| N_{sib_m}) =$$
$$\ldots$$
$$= H(H(...H(N_{par_0}' \| N_{sib_1})... \| N_{sib_{m-1}}) \| N_{sib_m}) =$$
$$= H(H(...H(H(D_i') \| N_{sib_1})... \| N_{sib_{m-1}}) \| N_{sib_m}).$$

The probability $P(R = R')$ is contingent on the cryptographic hash function $H$ used in the Merkle Tree. Assuming $H$ behaves as a random oracle, the probability of two different inputs producing the same hash output is negligible. Thus, we can express:

$$P(H(D_i) = H(D_i')) = \frac{1}{2^b} \quad (2)$$

where $b$ is the bit length of the hash output.

In blockchain networks, the integrity of transaction data is critical for maintaining trust and security. A low probability of data substitution ensures the reliability of the blockchain ledger. Similarly, in IoT ecosystems, where data from sensors and devices are continually aggregated, the integrity of this data is crucial for accurate analysis and decision-making. The ability to detect even minor alterations in data sets is vital for the overall security and functionality of these systems.

The primary objective of this article is to derive a method for the precise calculation of the probability (1) under the assumption (2). This probability (1) quantifies the likelihood of data falsification in a Merkle tree, thereby gauging the reliability of one of the most ubiquitous mechanisms for ensuring data integrity in blockchain technology and the Internet of Things. In the following sections, we present exact formulas for computing $P_{\text{falsification}}$, as well as approximate expressions that yield an exceptionally close approximation of this probability. Furthermore, we include the results of empirical experiments that corroborate the validity of these derived expressions.

## IV. THEORETICAL ESTIMATION OF FALSIFICATION PROBABILITY

The derivation of an exact formula to estimate the probability of data falsification (1) was conducted in a stepwise manner, considering various scenarios with different lengths of the Merkle path, denoted as $\mathcal{P}(D_i) = \{N_{sib_1}, N_{sib_2}, ..., N_{sib_m}\}$, for diverse values of $m$.

We commenced with the case of $m=0$, signifying that

$$P_{\text{falsification}} = P(R = R') = P(N_{par_0} = N_{par_0}'),$$

where

$$N_{par_0} = L_t = H(D_i), \quad N_{par_0}' = L_t' = H(D_i').$$

Consequently,

$$P_{\text{falsification}} = P(H(D_i) = H(D_i')) = \frac{1}{2^b}.$$

Thus, for $m=0$, the probability value (1) aligns with expression (2).

Considering the case of $m=1$, we have:

$$P_{\text{falsification}} = P(R = R') = P(N_{par_1} = N_{par_1}'),$$

where

$$N_{par_1} = H(N_{par_0} \| N_{sib_1}), \quad N_{par_0} = L_t = H(D_i),$$

and

$$N_{par_1}' = H(N_{par_0}' \| N_{sib_1}), \quad N_{par_0}' = L_t' = H(D_i').$$

The event $(N_{par_1} = N_{par_1}')$ occurs when corresponding hash codes match, i.e., when

$$H(N_{par_0} \| N_{sib_1}) = H(N_{par_0}' \| N_{sib_1}).$$

This equality is guaranteed when $N_{par_0} = N_{par_0}'$. However, this is only possible when the hash codes $H(D_i) = H(D_i')$ coincide. The probability of such an event, as previously shown, equals $\frac{1}{2^b}$.

In the scenario where $N_{par_0} \neq N_{par_0}'$, i.e., when for $D_i \neq D_i'$ the corresponding hash codes do not match: $H(D_i) \neq H(D_i')$), this case will be observed with the inverse probability $\left(1 - \frac{1}{2^b}\right)$.

Then, the coincidence $H(N_{par_0} \| N_{sib_1}) = H(N_{par_0}' \| N_{sib_1})$ will be observed with the probability $\left(1 - \frac{1}{2^b}\right)\frac{1}{2^b}$.

The final expression for calculating the probability (1) for $m=1$ becomes:

$$P_{\text{falsification}} = P\big(H(D_i) = H(D_i')\big) +$$
$$+ P\big(H(D_i) \neq H(D_i')\big) \cdot P\big(H(N_{par_0} \| N_{sib_1}) = H(N_{par_0}' \| N_{sib_1})\big) =$$
$$= \frac{1}{2^b} + \left(1 - \frac{1}{2^b}\right)\frac{1}{2^b}.$$

Extending similar reasoning to the case of $m=2$, we obtain:

$$P_{\text{falsification}} = P\big(H(D_i) = H(D_i')\big) +$$
$$+ P\big(H(D_i) \neq H(D_i')\big) \times P\big(H(N_{par_0} \| N_{sib_1}) = H(N_{par_0}' \| N_{sib_1})\big) +$$
$$+ P\big(H(D_i) \neq H(D_i')\big) \times P\big(H(N_{par_0} \| N_{sib_1}) \neq H(N_{par_0}' \| N_{sib_1})\big) \times$$
$$\times P\big(H(H(N_{par_0} \| N_{sib_1}) \| N_{sib_2}) = H(H(N_{par_0}' \| N_{sib_1}) \| N_{sib_2})\big) =$$
$$= \frac{1}{2^b} + \left(1 - \frac{1}{2^b}\right)\frac{1}{2^b} + \left(1 - \frac{1}{2^b}\right)\left(1 - \frac{1}{2^b}\right)\frac{1}{2^b}.$$

Generalizing this formula for any positive integer $m$, we derive the general formula:

$$P_{\text{falsification}} = \frac{1}{2^b} + \left(1 - \frac{1}{2^b}\right)\frac{1}{2^b} + \left(1 - \frac{1}{2^b}\right)\left(1 - \frac{1}{2^b}\right)\frac{1}{2^b} + ... +$$
$$+ \underbrace{\left(1 - \frac{1}{2^b}\right)\left(1 - \frac{1}{2^b}\right)...\left(1 - \frac{1}{2^b}\right)}_{m \text{ times}}\frac{1}{2^b} = \frac{1}{2^b} + \sum_{k=1}^{m}\left(1 - \frac{1}{2^b}\right)^k \frac{1}{2^b}.$$

The sum on the right side of the last expression can be simplified using the rule for calculating the sum of the first $m$ terms of a geometric progression.

Let $g$ be the first term of the geometric progression, and $z$ the common ratio, i.e., the factor by which each term is multiplied to obtain the next one. Then, the formula for the sum of the first $m$ terms of a geometric progression is:

$$G_m = g \cdot \frac{1 - z^m}{1 - z}.$$

In our case, $g = z = 1 - \frac{1}{2^b}$, thus:

$$G_m = \sum_{k=1}^{m}\left(1-\frac{1}{2^b}\right)^k =$$

$$= \left(1-\frac{1}{2^b}\right)\frac{1-\left(1-\frac{1}{2^b}\right)^m}{1-\left(1-\frac{1}{2^b}\right)} = \left(-1+2^b\right)\left(1-\left(1-\frac{1}{2^b}\right)^m\right).$$

Substituting $G_m$ into the formula for $P_{\text{falsification}}$, we obtain:

$$P_{\text{falsification}} = \frac{1}{2^b} + \sum_{k=1}^{m}\left(1-\frac{1}{2^b}\right)^k \frac{1}{2^b} =$$

$$= \frac{1}{2^b} + \frac{1}{2^b}\left(-1+2^b\right)\left(1-\left(1-\frac{1}{2^b}\right)^m\right) = \quad (3)$$

$$= \frac{1}{2^b} + \left(1-\frac{1}{2^b}\right)\left(1-\left(1-\frac{1}{2^b}\right)^m\right).$$

The derived formula (3) allows for an exact calculation of the probability of data falsification when using the Merkle path $\mathcal{P}(D_i) = \{N_{sib_1}, N_{sib_2}, ..., N_{sib_m}\}$ and under the assumption of truth in (2). As evident from formula (3), the probability $P_{\text{falsification}}$ increases as $m$, the number of elements in the Merkle path $\mathcal{P}(D_i)$, increases.

To derive an approximate formula, we note that when $b$ takes large values, the magnitude of $\frac{1}{2^b}$ becomes very small. This allows for an approximation using the initial terms of the Taylor series for $\exp(x)$, where $x = \frac{-1}{2^b}$. Consequently, the approximation can be expressed as: $\exp\left(\frac{-1}{2^b}\right) \approx 1 - \frac{1}{2^b}$.

Substituting this approximation into formula (3), we obtain:

$$P_{\text{falsification}} \approx \frac{1}{2^b} + \exp\left(\frac{-1}{2^b}\right) - \exp\left(\frac{-m-1}{2^b}\right). \quad (4)$$

V. VERIFICATION OF THEORETICAL FORMULAS THROUGH EMPIRICAL TESTING

In the experimental phase of our research, we focused on empirically validating the theoretical formula (3) derived in the previous sections. To achieve this, we developed a Python program, as referenced in [15], designed to estimate the probability of data falsification and compare it with our theoretical calculations.

Our program implementation involved several key functions and methodologies:
- Hash Generation: We utilized the SHA256 hashing algorithm, implemented in Python's hashlib module, to generate hashes of data. The generate_hash function takes a data string and returns a truncated hash of specified length.
- Random Data Generation: The generate_random_data function creates random strings of a specified length, combining ASCII letters and digits. This function is crucial for simulating various data inputs in the experiment.
- Merkle Root Calculation: The calculate_merkle_root function computes the root hash for given data and a Merkle path. It iteratively hashes the data with each element of the path, simulating the process of ascending a Merkle Tree.
- Theoretical Probability Calculations: We implemented two different theoretical probability functions (3) and (4) to provide a comprehensive theoretical framework for comparison with empirical results.
- Experiment Execution: The run_experiment function conducts the empirical testing. It generates a random Merkle path, calculates the root hash, and then counts the number of matches found when recalculating the root hash with new random data. This process is repeated across a specified number of experiments and iterations.
- Graphical Representation: We used matplotlib to plot the results. The empirical probabilities of data falsification for different hash lengths and Merkle path sizes were plotted alongside the theoretical estimates. This visual representation aids in directly comparing empirical data with theoretical predictions.

In this study, a comprehensive experimental approach was adopted to rigorously assess the accuracy of the approximation formula in comparison to the exact formula. For each set of parameters $b$ and $m$, a total of 1000 individual experiments were conducted, and this process was repeated 100 times. This methodology was meticulously designed to amass a substantial volume of data, thereby facilitating a robust statistical analysis.

The rationale behind this extensive experimental repetition lies in its ability to mitigate the impact of random variations and anomalies, ensuring the reliability and validity of the results. By aggregating data from 100,000 experiments for each parameter set, the study aimed to achieve a high level of precision in its findings, thereby providing a solid foundation for drawing statistically significant conclusions.

Figure 1 presents the empirical results of our experiments, illustrating the relationship between the probability of data falsification and the variables $b$ (hash length) and $m$ (number of elements in the Merkle path).

The graph reveals two key trends:
1. Decrease in Falsification Probability with Increasing $b$: As the hash length $b$ increases, the probability of data falsification decreases significantly. This trend is consistent with cryptographic principles, where longer hash lengths correspond to a larger space of possible hash values, thereby reducing the likelihood of hash collisions. In the context of Merkle Trees, a longer hash implies a lower probability that two different data inputs will yield the same Merkle Root, enhancing the security and integrity of the data.
2. Increase in Falsification Probability with Increasing $m$: Conversely, the graph shows an increase in the probability of data falsification with an increasing number of elements $m$ in the Merkle path. This observation can be attributed to the cumulative effect of hash collisions along the Merkle path. As the number of elements in the path increases, the probability of encountering a hash collision at some point in the path also increases, even if each individual hash operation remains secure. This effect highlights a trade-off in Merkle Tree design: while a longer path provides a more detailed verification trail, it also slightly increases the overall risk of hash collisions.

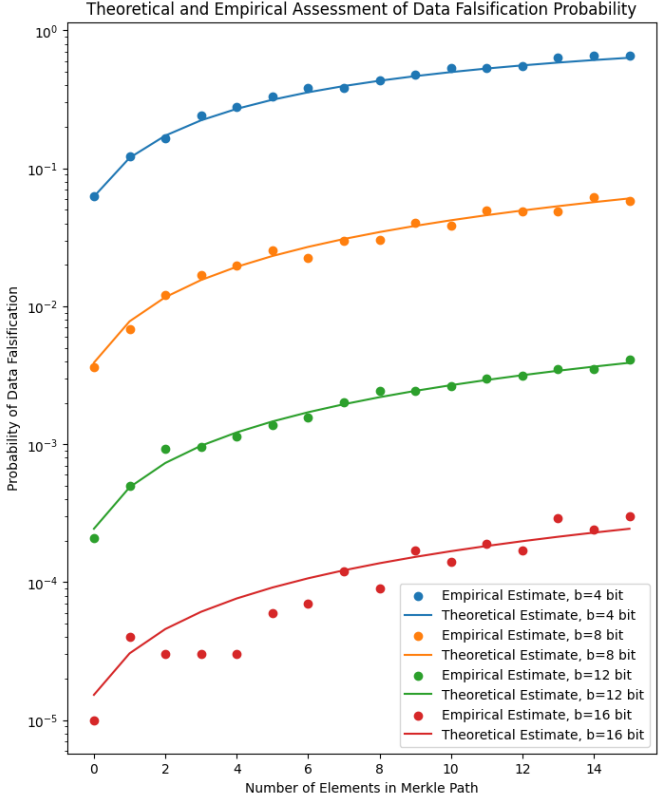

**Fig. 1.** Theoretical and Empirical Assessment of Data Falsification Probability

The empirical results align well with the theoretical predictions, confirming the validity of our theoretical models. The graph serves as a crucial tool for understanding the dynamics of data integrity in systems employing Merkle Trees. For blockchain and IoT applications, where data security is paramount, these findings underscore the importance of optimizing hash length and Merkle path length to balance security and computational efficiency.

In the second part of our experimental analysis, we focused on quantifying the discrepancy between the exact and approximate probabilities of data falsification in Merkle Trees. This investigation aimed to validate the precision of our approximate formula (4) against the exact calculations (3), particularly under varying conditions of hash length $b$ and Merkle path length $m$. The numerical comparison involved calculating the absolute difference between the values obtained using both formulas across various combinations of the parameters $b$ and $m$. The results of the experiment are presented in the Table I.

TABLE 1. ABSOLUTE DIFFERENCE BETWEEN THE VALUES OBTAINED USING FORMULAS (3) AND (4)

| $b$ | $m$ | Difference |
|---|---|---|
| 2 | 10 | 0.00710805789680180 |
| 2 | 50 | 0.0287983054922386 |
| 2 | 100 | 0.0288007830608296 |
| 2 | 500 | 0.0288007830714048 |
| 2 | 1000 | 0.0288007830714048 |
| 4 | 10 | 0.009236819799283665 |
| 4 | 50 | 0.00216253840990199 |
| 4 | 100 | 0.00157561166419240 |
| 4 | 500 | 0.00191306281345960 |
| 4 | 1000 | 0.00191306281347581 |
| 6 | 10 | 0.00102043490152098 |
| 6 | 50 | 0.00270533021104558 |
| 6 | 100 | 0.00243368381652498 |
| 6 | 500 | 0.0000975622497489947 |
| 6 | 1000 | 0.000121418280353613 |
| 8 | 10 | 0.0000729807479756192 |
| 8 | 50 | 0.000311967273480596 |
| 8 | 100 | 0.000512896153700371 |
| 8 | 500 | 0.000532762483821725 |
| 8 | 1000 | 0.000145220188735085 |
| 10 | 10 | 0.00000471585051471136 |
| 10 | 50 | 0.00002267528748744780 |
| 10 | 100 | 0.0000431877859932567 |
| 10 | 500 | 0.000146063524593065 |
| 10 | 1000 | 0.000179180050212557 |

The results of our numerical analysis revealed several key insights:
- For smaller values of $b$ (e.g., 2 and 4), the discrepancy between the exact and approximate probabilities was more pronounced, especially for larger values of $m$. This indicates that the approximation may be less accurate for shorter hash lengths and longer Merkle paths.
- As $b$ increased, the discrepancy consistently decreased, highlighting the improved accuracy of the approximate formula for longer hash lengths. For instance, at $b=10$, the differences were minimal across all tested values of m.

These findings are significant for practical applications in blockchain and IoT systems. The ability to accurately approximate the probability of data falsification is crucial for assessing the security and integrity of these systems.

## VI. DISCUSSION

This section delves into the synthesis of our theoretical findings and experimental outcomes, aiming to provide a comprehensive understanding of the implications and applications of our research in the realms of blockchain and the Internet of Things (IoT).
- Theoretical Insights: Our theoretical exploration into the probability of data falsification in Merkle Trees yielded significant insights. The derived formulas, both exact and approximate, offer a mathematical framework to assess the integrity of data within these structures. The theoretical models underscore the critical role of hash length ($b$) and Merkle path length ($m$) in determining the security of the data. Particularly, the inverse relationship between hash length and falsification probability highlights the importance of employing sufficiently long hashes in practical applications to minimize security risks.

- Empirical Validation: The empirical experiments conducted provided a robust validation of our theoretical models. The observed decrease in falsification probability with increasing hash length and the converse effect with increasing Merkle path length align well with our theoretical predictions. These results not only reinforce the reliability of the theoretical models but also demonstrate their practical applicability in real-world scenarios.
- Numerical Analysis: The numerical analysis comparing the exact and approximate probabilities of data falsification further enriched our understanding. The findings revealed that while the approximate formula generally holds well, especially for longer hash lengths, its accuracy diminishes for shorter hash lengths and longer Merkle paths. This insight is crucial for practitioners in blockchain and IoT, suggesting a careful evaluation of these parameters when implementing Merkle Trees.
- Practical Implications: From a practical standpoint, our research offers valuable guidelines for optimizing the security and efficiency of blockchain and IoT systems. The balance between hash length and Merkle path length is a key consideration for system designers, as it directly impacts the probability of data falsification and, consequently, the overall system integrity. Our findings suggest that while longer hashes enhance security, they also entail computational overhead, necessitating a judicious choice based on the specific requirements of the application.
- Future Research Directions: The complexity observed in the interaction between hash length and Merkle path length opens avenues for further research. Investigating the optimal combinations of these parameters for different application scenarios could lead to more refined guidelines for system design. Additionally, exploring the impact of different hashing algorithms on the probability of data falsification could provide deeper insights into the security aspects of Merkle Trees.

## VII. CONCLUSIONS

In conclusion, our comprehensive analysis blending theoretical models with empirical and numerical validations offers a nuanced understanding of data integrity in Merkle Trees. The insights gained from this study are instrumental in advancing the security frameworks of blockchain and IoT systems, contributing significantly to the field of data integrity and security.

ACKNOWLEDGMENT

This project is supported by Proxima Labs, 1501 Larkin Street, suite 300, San Francisco, USA.